\documentclass[aps,pre,twocolumn,superscriptaddress]{revtex4-2}
\usepackage{graphicx}
\usepackage{xcolor}
\usepackage{dcolumn}
\usepackage{bm}
\usepackage{amsmath}
\usepackage{amssymb}

\begin{document}

\title{Coarsening dynamics for spiral and disordered waves in active Potts models}

\author{Hiroshi Noguchi}
\email[]{noguchi@issp.u-tokyo.ac.jp}
\affiliation{Institute for Solid State Physics, University of Tokyo, Kashiwa, Chiba 277-8581, Japan}
\date{\today}

\begin{abstract}
This study examines the domain-growth dynamics of $q$-state active Potts models ($q=3$--$8$) under the cyclically symmetric conditions using Monte Carlo simulations on square and hexagonal lattices.
By imposing active cyclic flipping of states, finite-length waves emerge in the long-term limit. 
This study focuses on coarsening dynamics from an initially random mixture of states to these moving-domain states.
The correlation length and mean cluster size grow, following the Lifshitz--Allen--Cahn (LAC) law ($\propto t^{1/2}$) in the intermediate time range,
and in the late range, saturation is observed at the characteristic wavelengths.
It is found that the  growth rate is raised prior to saturation, leading to a transient increase in the coarsening exponent. 
The coarsening dynamics to disordered waves exhibit greater transient increases than those to spiral waves.
Moreover, the transient increase is greater at higher $q$.
In factorized symmetry modes at $q=6$, domains composed of two or three states similarly follow the LAC law.
Finally, this study confirms that the choice of lattice type (square or hexagonal) and update scheme (Metropolis or Glauber) does not alter the dynamic behavior.
\end{abstract}

\maketitle

\section{Introduction}

Domain coarsening is a well-known phenomenon observed during spinodal decomposition in rapidly quenched systems,
and its dynamics are theoretically  well established~\cite{puri09,furu85,bray94,gres88,huse86,roge88,tana97,nogu06a,ferr07,midy20,yang25,yuan25}.
In particular, for surface-tension-driving coarsening in a nonconserved order parameter,
the domain length grows proportionally to $t^{1/2}$, following the Lifshitz--Allen--Cahn (LAC) law~\cite{lifs62,alle79}. By contrast, diffusion-dominated conserved systems exhibit the growth in accordance with the  Lifshitz--Slyozov law ($t^{1/3}$)~\cite{lifs61}.
However, coarsening dynamics leading to nonequilibrium states have been far less explored~\cite{oppo87,hamm05,vazq08,lato22,elde92,cros95,dey12,mish14,saha20,katy20,patt21,ditt23,band24,rouz25}.
Under nonequilibrium conditions, the final states can be nonstatic and manifest as spatiotemporal patterns, such as traveling waves and global oscillations~\cite{nico77,hake04,mikh94,rabi00,kura84,murr03,beta17,bail22,nogu24c}.
Recently, increasing attention has been directed toward the coarsening to such nonstatic states~\cite{elde92,cros95,dey12,mish14,saha20,katy20,patt21,ditt23,band24,rouz25}.
In conserved systems, self-propelling activity has been shown to induce coarsening exponents either higher~\cite{mish14} or lower~\cite{saha20,katy20,patt21} 
than $1/3$. By contrast, coarsening in nonconserved active systems has been hardly studied.

The study aims to understand the coarsening dynamics to spatiotemporal patterns in a nonconserved system by using Monte Carlo (MC) simulations in two-dimensional (2D) lattices.
We employ a $q$-state Potts model extended to a nonequilibrium situation by imposing cyclic flipping of states~\cite{nogu24a,nogu24b,nogu25,nogu25b,nogu25a,nogu26a}. Since the coarsening dynamics to equilibrium states have been extensively studied using lattice models~\cite{puri09,furu85,bray94,gres88},
the influence of nonequilibrium activity can be easily identified.
In these Potts models, several types of stable spatiotemporal patterns can be obtained in the long-term limit.
We examine four types of final spatiotemporal patterns by adjusting parameters:
spiral and disordered waves of single-state dominant domains of $q$ states at $q=3$--$8$, and
spiral waves of mixed domains and coexistence of two spiral-wave phases at $q=6$.

The models and methods are described in Sec.~\ref{sec:model}.
Simulation results are presented and discussed in Sec.~\ref{sec:results}.
The results of Potts models on square lattices are described in Secs.~\ref{sec:3}--\ref{sec:q6t1}.
The coarsening dynamics at $q=3$ are presented in Sec.~\ref{sec:3}.
The results of standard and repulsive nonflip-contact Potts models at $q=4$--$8$ are presented in Secs.~\ref{sec:d0} and \ref{sec:d2}, respectively.
The results of attractive and repulsive diagonal-contact Potts models at $q=6$
are presented in Secs.~\ref{sec:q6tn15} and \ref{sec:q6t1}, respectively.
The robustness of the simulation dynamics is discussed in Sec.~\ref{sec:tri}, where
 hexagonal lattice and Glauber MC update schemes are examined.
Finally, a summary is presented in Sec.~\ref{sec:sum}.

\section{Models and Methods}\label{sec:model}

For the 2D square and hexagonal lattices, 
square and rhombus simulation cells with a side length of $L$ are employed, respectively, under periodic boundary conditions.
The total number of sites is $N=L^2$.
In the $q$-state Potts model, each site possesses a state $s\in [0,q-1]$.
The nearest neighboring sites ($i$ and $j$) interact with contact energies $J_{s_i,s_j}$:~\cite{wu82,pott52}
\begin{equation}
\label{eq:hint}
H_{\mathrm{int}} = - \sum_{\langle ij\rangle} J_{s_is_j}.
\end{equation}
In the square and hexagonal lattices, each site has four and six nearest neighbors, respectively.
In equilibrium systems, each state may additionally possess self-energy $\varepsilon_s$,
and the ratio of the forward and backward flip rates for  flipping of a single site from $s$ to $s'$ is given by $\exp(-\Delta H_{s_is'_i})$, where $\Delta H_{s_is'_i} = \Delta H_{\mathrm{int}} - h_{s,s'}$ is the energy difference between the two states, 
$h_{s,s'} = \varepsilon_s - \varepsilon_s'$ is the flipping energy,
and the thermal energy $k_{\mathrm{B}}T$ is normalized to unity.
At equilibrium, the cyclic sum of the flipping energies is $\sum_{k}^q h_{k,[k+1]} = 0$, where $[k]$ denotes $k \bmod q$.
This model is extended to a nonequilibrium framework, where $\sum_{k}^q h_{k,[k+1]} > 0$, while maintaining $h_{ss'}=-h_{s's}$~\cite{nogu24a}.
Therefore, the detailed balance can be locally satisfied for flips between $s$ and $s'$, but not globally over a full cycle ($s=0 \to 1\ ... \to (q-1) \to 0$). For $q=3$, this corresponds to the rock--paper--scissors relationship.
Such dynamics can be realized experimentally~\cite{nogu24a,nogu25b,nogu25a}, for example, in reactions on a catalytic surface~\cite{ertl08,bar94,goro94,barr20,zein22} and molecular transport across membranes~\cite{tabe03,miel20,holl21,nogu23}, in which energy is input as a chemical reaction and as a chemical potential difference between two solutions, respectively.

In this study, the analysis is restricted to cyclically symmetric  conditions: constant flipping energy $h_{k,[k+1]}=h$, and $J_{k,k'}$ depends solely on the state distance $|k-k'|$ (see Fig.~S1 in the Supplemental Material~\cite{SI}). For example, a six-state Potts model has three independent contact energy parameters: $J_{k,k}-J_{k,[k+1]}$, $J_{k,[k+2]}-J_{k,[k+1]}$, and $J_{k,[k+3]}-J_{k,[k+1]}$.
A site is randomly selected, and
its flip to a neighboring state ($s=k \to [k+1]$ or $[k-1]$) is performed using the Metropolis MC algorithm with the acceptance ratio calculated as $p_{s_is'_i}=\min[1, \exp(-\Delta H_{s_is'_i})]$.
This flip is attempted $N$ times per MC step (time unit).
We previously verified that the choice between the Metropolis and Glauber schemes for the MC update introduces only a minor difference in the long-term dynamics at $q=3$~\cite{nogu24a}.
In the Glauber scheme, the acceptance ratio $p_{s_is'_i}=1/[1+\exp(\Delta H_{s_is'_i})]$ is used in the MC update.
In Sec.~\ref{sec:tri}, we further discuss the effects of the update schemes on the coarsening dynamics at $q=3$ and $6$.

In this study,
we set $J_{k,k}=2$ and $1.2$ for the square and hexagonal lattices, respectively, and  $J_{k,[k+1]}=0$ for both lattices. 
We use the standard Potts models ($J_{k,[k+n]}= 0$)
and repulsive nonflip-contact Potts models ($J_{k,[k+n]} =-J_{k,k}$) with $2\le n\le q/2$
to examine $q$ dependence.
Note that these two models are identical at $q=3$.
When $q$ is a factorizable number, modes with factorized symmetry can emerge~\cite{nogu25b}.
As representative modes, the coarsening dynamics to two wave modes are simulated with factorized symmetry at  $q=6$.
i) M2W3 mode: three types of mixed domains form spiral waves under conditions $J_{k,[k+2]} =0$ and $J_{k,[k+3]} \sim 2$.
The neighboring pairs of diagonal states ($s=k$ and $[k+3]$) 
have stronger attractive interactions than the other pairs of different states, resulting in a mixed domain.
Here, we use $J_{k,[k+3]} =1.5$ and refer to this as the attractive diagonal-contact condition.
ii) W3 mode: three odd- or even-numbered states form spiral waves under conditions $J_{k,[k+2]} =0$ and $J_{k,[k+3]} \simeq -1$.
These two spiral waves temporally coexist through stochastic switching in the long term,
whereas they coexist spatially during coarsening.
Here, we use $J_{k,[k+3]} = -1$ and refer to this as the repulsive diagonal-contact condition.

\begin{figure}[tbh]
\includegraphics[]{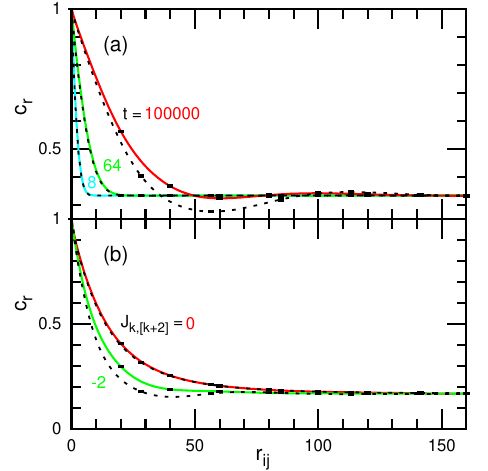}
\caption{
Spatial correlation function $C_{\mathrm{r}}(r,t)$.
(a) Three-state Potts model at $t=8$, $64$, and $100~000$ for $h=0.5$.
(b) Six-state standard ($J_{k,[k+2]}=J_{k,[k+3]}=0$) and repulsive nonflip-contact ($J_{k,[k+2]}=J_{k,[k+3]}=-2$) Potts models  at  $t=100~000$.
The solid and dashed lines represent the data along the axial [$(x,y)=(1, 0)$ and $(0, 1)$] and diagonal [$(x,y)=(1, \pm 1)$] directions, respectively.
The error bars are shown at several data points, but most of them are smaller than the line thickness.
}
\label{fig:acf}
\end{figure}

Domain lengths are calculated from the spatial correlation and mean cluster size (area).
The spatial correlation function $C_{\mathrm{r}}({\mathbf{r}},t) = (1/N)\sum_{i,j}\delta_{s_i,s_j}\delta({\mathbf{r}(t)}_{i,j}-{\mathbf{r}(t)})$ is 
calculated along the axial [$(x,y)=(1, 0)$ and $(0, 1)$] and diagonal [$(x,y)=(1, \pm 1)$] directions for the square lattices (see Fig.~\ref{fig:acf}).
Three directions [$(x,y)= (1,0), (1/2,\pm\sqrt{3}/2)$] along the three sides of triangles are used for the hexagonal lattices.
The half-relaxation lengths along the axial directions are considered as the correlation lengths.
For single-state dominant domains,
the half-relaxation length $r_{\mathrm{cr1}}(t)$ is calculated as $C_{\mathrm{r}}(r_{\mathrm{cr1}},t)=(1+1/q)/2$, 
since $C_{\mathrm{r}}(r,t)=1/q$ at $r\to\infty$.
For domains comprising $n$ states,
$r_{\mathrm{cr}n}(t)$ is calculated as $C_{\mathrm{r}}(r_{\mathrm{cr}n},t)=(1+n/q)/2$.
The neighboring sites occupied by the same state belong to the same cluster.
The mean cluster size of the $s=k$ state is given by 
$n_{k,\mathrm {cl}}= (\sum_{i_{k,{\mathrm {cl}}}=1}^{N_k} {i_{k,\mathrm{cl}}}^2 n^{k,{\mathrm {cl}}}_i)/N_k$,
where $n^{k,{\mathrm {cl}}}_i$  is the number of clusters of size $i_{k,\mathrm{cl}}$,  and
$N_k=i_{k,\mathrm{cl}} n^{k,{\mathrm {cl}}}_i$ is the total number of $s=k$ sites.
Cluster sizes are averaged as $n_{\mathrm {cl}}=\langle n_{k,\mathrm {cl}}\rangle$,
where $\langle ... \rangle$ is the ensemble average for the $q$ states and different simulation runs.
The characteristic length based on the domain area is given by $n_{\mathrm {cl}}^{1/2}$.
We primarily use $L=2048$ ($N=4~194~304$). To check the finite size effects,  $L=1024$ is also used at several parameter sets.
The coarsening exponent $\alpha_{\mathrm{cr}}(t)$ is calculated as~\cite{huse86}
\begin{equation}
  \alpha_{\mathrm{cr}}(t) = \log_b[r_{\mathrm{cr1}}(bt)/r_{\mathrm{cr1}}(t)],
\end{equation}
with $b=4$. Similarly, $\alpha_{\mathrm{cl}}(t)$ for $n_{\mathrm {cl}}^{1/2}$.
The averages of the exponents for short time ranges are represented by $\bar{\alpha}$.
We also calculate the contact probability $p_{\mathrm{cn}}$ of different states in neighboring sites.
For $q=3$, an $s=0$ site has $p_{\mathrm{cn}}=1/4$ when the state of one neighbor site is $s=1$ and the others are $s=0$ in a square lattice.
Then the probabilities are averaged for all sites.
For $q>3$, $p_{\mathrm{cn1}}, p_{\mathrm{cn2}}, \dotsc, p_{\mathrm{cn}(q/2)_{\mathrm{int}}}$ are the contact probabilities of the first, second, $\dotsc$, $(q/2)_{\mathrm{int}}$-th neighboring states, respectively,
where $(q/2)_{\mathrm{int}}$ is he largest integral value that is not greater than $q/2$.
Statistical errors are calculated from ten independent runs.

\begin{figure}[tbh]
\includegraphics[]{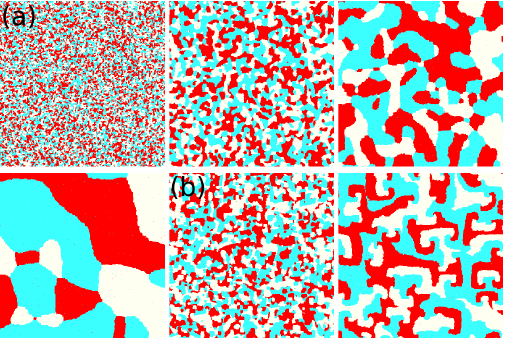}
\includegraphics[]{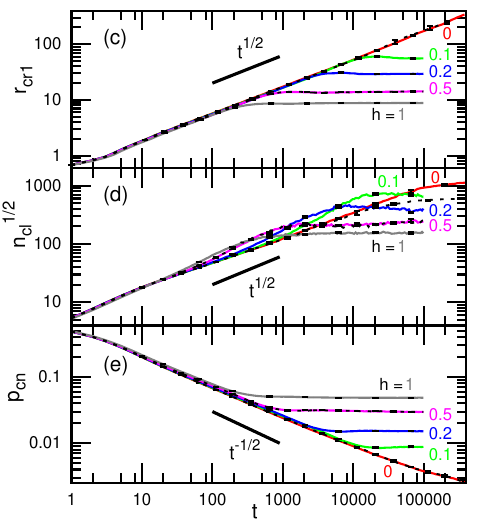}
\caption{
Coarsening dynamics of the three-state standard Potts models.
(a)--(b) Sequential snapshots of a local region ($L/4\times L/4$) at $L=2048$.
(a) Snapshots at $t=10$, $100$, $1000$, and $10000$ for $h=0$.
(b) Snapshots at $t=100$ and $10000$ for $h=0.5$.
The light yellow, cyan, and red sites (light to dark in grayscale) 
represent $s=0$, $1$, and $2$, respectively.
(c)--(e) Time evolution of 
(c) the correlation length $r_{\mathrm{cr1}}$, (d) root of the mean cluster size $n_{\mathrm{cl}}^{1/2}$, and 
(e) contact probability $p_{\mathrm{cn}}$ of different states.
The solid lines represent the data for $h=0$, $0.1$, $0.2$, $0.5$, and $1$ at $L=2048$.
The black dashed lines represent the data for $h=0$ and $0.5$ at $L=1024$.
The short black lines in (c)--(e) represent the slopes of $t^{1/2}$,  $t^{1/2}$, and  $t^{-1/2}$, respectively.
The error bars are shown at several data points, but most of them are smaller than the line thickness. 
}
\label{fig:q3}
\end{figure}

\begin{figure}[tbh]
\includegraphics[]{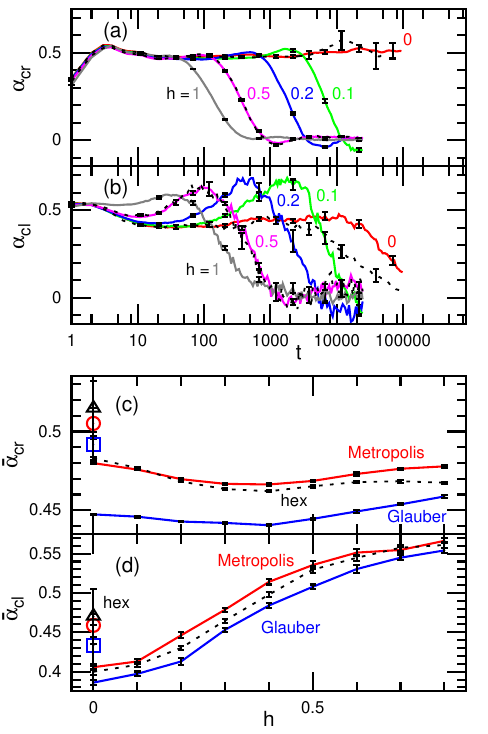}
\caption{
Coarsening exponents $\alpha_{\mathrm{cr}}$ and $\alpha_{\mathrm{cl}}$ for $q=3$.
(a)--(b) Time evolution of (a) $\alpha_{\mathrm{cr}}$ and (b) $\alpha_{\mathrm{cl}}$.
(c)--(d) Local averages $\bar{\alpha}_{\mathrm{cr}}$ and $\bar{\alpha}_{\mathrm{cl}}$ as functions of $h$.
The red and blue solid lines represent the data obtained using the Metropolis and Glauber update schemes, respectively, on the square lattice.
The black dashed lines represent the data obtained using the Metropolis scheme on the hexagonal lattice.
The solid lines represent the averages for $2<r_{\mathrm{cr1}}(t)<4$ and for $20<n_{\mathrm{cl}}(t)^{1/2}<50$.
The circles, squares, and triangles  represent the averages for $30<r_{\mathrm{cr1}}(t)<80$ and 
for $200<n_{\mathrm{cl}}(t)^{1/2}<300$ at $h=0$.
}
\label{fig:slope_q3}
\end{figure}

\begin{figure}[tbh]
\includegraphics[]{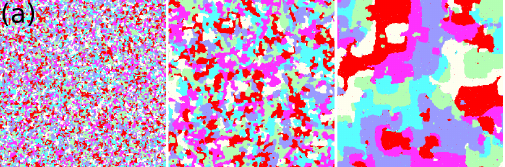}
\includegraphics[]{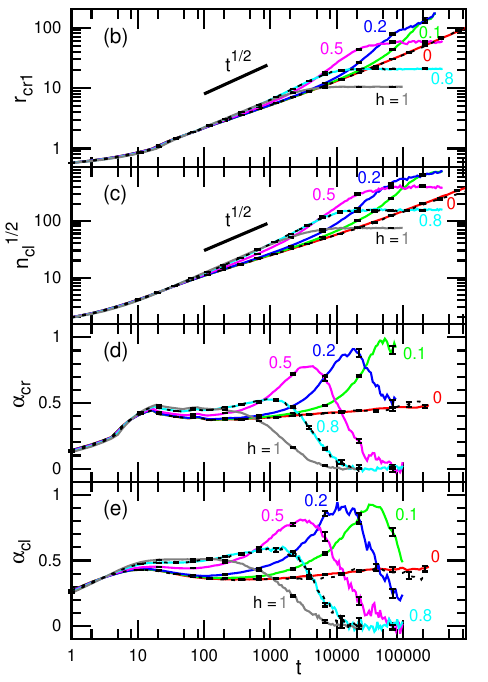}
\caption{
Coarsening dynamics of the six-state standard Potts model ($J_{k,[k+2]}=J_{k,[k+3]}=0$).
(a) Sequential snapshots ($L/4\times L/4$) at $t=100$, $1000$, and $100~000$ for $h=0.8$ and $L=2048$.
The light yellow, light green, cyan, blue, magenta, and red sites (light to dark in grayscale) 
represent $s=0$, $1$, $2$, $3$, $4$, and $5$, respectively.
(b)--(e) Time evolution of (b) the correlation length $r_{\mathrm{cr1}}$, (c) root of the mean cluster size $n_{\mathrm{cl}}^{1/2}$,
(d) exponents $\alpha_{\mathrm{cr}}$, and (e) $\alpha_{\mathrm{cl}}$.
The solid lines represent the data for $h=0$, $0.1$, $0.2$, $0.5$, $0.8$, and $1$ at $L=2048$.
The black dashed lines represent the data for $h=0$ and $0.8$ at $L=1024$.
The short black lines in (b) and (c) represent the slope of $t^{1/2}$.
The error bars are shown at several data points.
}
\label{fig:q6d0}
\end{figure}

\begin{figure}[t]
\includegraphics[]{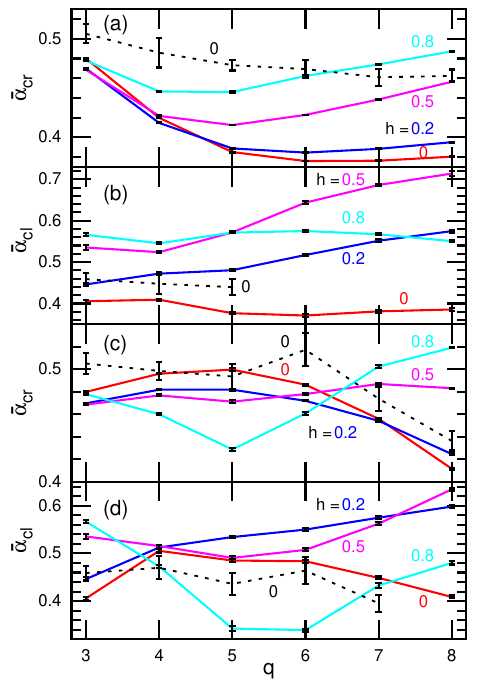}
\caption{
Local averages $\bar{\alpha}_{\mathrm{cr}}$ and $\bar{\alpha}_{\mathrm{cl}}$ of coarsening exponents  as functions of $q$
at  $h=0$, $0.2$, $0.5$, and $0.8$.
(a)--(b) Standard Potts models.
(c)--(d) Repulsive nonflip-contact Potts models.
The solid lines represent the averages for $2<r_{\mathrm{cr1}}(t)<4$ and for $20<n_{\mathrm{cl}}(t)^{1/2}<50$.
The dashed lines represent the averages for $30<r_{\mathrm{cr1}}(t)<80$ and 
for $200<n_{\mathrm{cl}}(t)^{1/2}<300$ at $h=0$.
}
\label{fig:slope}
\end{figure}

\begin{figure}[t]
\includegraphics[]{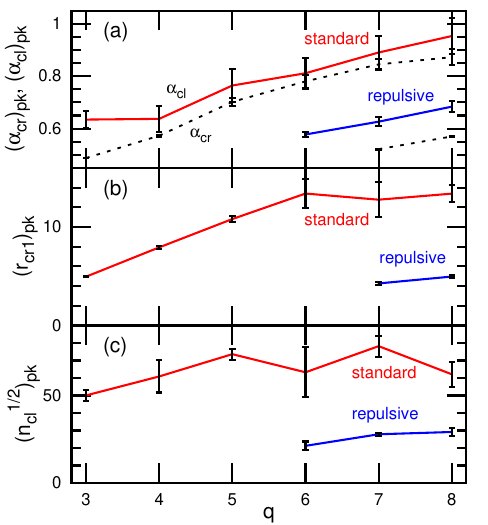}
\caption{
Peaks in coarsening exponents $\alpha_{\mathrm{cr}}$ and $\alpha_{\mathrm{cl}}$ as functions of $q$ at  $h=0.5$
for the standard and repulsive nonflip-contact Potts models.
(a) Peak heights of the exponents.
The solid and dashed lines represent the data of $\alpha_{\mathrm{cl}}$ and $\alpha_{\mathrm{cr}}$, respectively.
(b) Correlation length $r_{\mathrm{cr1}}$ at the peaks in $\alpha_{\mathrm{cr}}$.
(c) Cluster length $n_{\mathrm{cl}}^{1/2}$ at the peaks in $\alpha_{\mathrm{cl}}$.
}
\label{fig:peak}
\end{figure}

\section{Results and Discussion}\label{sec:results}

\subsection{Three-State Potts Model}\label{sec:3}

It is well established that the coarsening dynamics to the thermal equilibrium state ($h=0$) at $q=3$
follow the LAC law with the characteristic length scaling as $r \propto t^{1/2}$~\cite{bray94,gres88}.
Our simulations confirm this dynamic law, as shown in Fig.~\ref{fig:q3}.
Both the correlation length $r_{\mathrm{cr1}}$ and cluster length $n_{\mathrm{cl}}^{1/2}$ increase as $t^{1/2}$.
The contact probability $p_{\mathrm{cn}}$ of different states (proportional to domain boundary length $r_{\mathrm{bd}}$)
follows as $p_{\mathrm{cn}}\propto t^{-1/2}$, because $r_{\mathrm{bd}} \propto N/r_{\mathrm{cr1}}$~\cite{ferr07}.
The cluster length  is saturated to $n_{\mathrm{cl}}^{1/2}\simeq 600$ and $1200$ at $L=1024$ and $2048$, respectively,
owing to the finite-size effect [see Fig.~\ref{fig:q3}(d)], whereas $r_{\mathrm{cr1}}$ does not exhibit such saturation 
within the simulated time range  [see Fig.~\ref{fig:q3}(c)].
Since the exponent $1/2$ is the asymptotic value for $t\to \infty$,
the exponent values obtained at finite times are slightly smaller
and  asymptotically approach $1/2$. [see Fig.~\ref{fig:slope_q3}].
Comparable deviations have been reported in previous simulations~\cite{huse86,roge88}.

At $h>0$, spiral waves emerge at $t\to \infty$, where the boundaries between the $s=k$ and $[k+1]$ domains move ballistically in a rock--paper--scissors manner [see Fig.~\ref{fig:q3}(b)].
The number density of spirals increases linearly with $h$~\cite{nogu24a}.
Consequently, $r_{\mathrm{cr1}}$,  $n_{\mathrm{cl}}^{1/2}$, and $p_{\mathrm{cn}}$ are saturated to finite values,
with saturation occurring earlier at higher $h$.
However, the temporal evolution of $r_{\mathrm{cr1}}$ and $p_{\mathrm{cn}}$ remains nearly identical to that at $h=0$ until
the onset of saturation [see Fig.~\ref{fig:q3}(c) and (e)].
Hence, the coarsening exponent in the range $r_{\mathrm{cr1}}\lesssim 4$ exhibits little dependence on $h$ 
[see Fig.~\ref{fig:slope_q3}(a) and (c)].

In contrast to $r_{\mathrm{cr1}}$ and $p_{\mathrm{cn}}$, 
$n_{\mathrm{cl}}^{1/2}$ exhibits more rapid growth prior to saturation [see Figs.~\ref{fig:q3}(d) and S2(a)], with the coarsening exponent transiently increasing with  $h$ [Fig.~\ref{fig:slope_q3}(b) and (d)].
This behavior arises from the elongated crescent shapes of the domains in spiral waves (compare the snapshots in Figs.~\ref{fig:q3}(a)--(b) at the same time steps).
During this transient regime, the domain morphology evolves from approximately circular to crescent-shaped.
It is rather interesting that $r_{\mathrm{cr1}}$ maintains its $t^{1/2}$ development in this time range.
Note that no system-size effects are detected at $h=0.5$, since the final domains are much smaller than the entire system [compare solid and dashed lines for $h=0.5$ in Fig.~\ref{fig:q3}(c)--(e)].

The domains in spiral waves have a rectangular shape elongating along the axial directions of the square lattice [$(x,y)=(1, 0)$ and $(0, 1)$], as shown in Fig.~\ref{fig:q3}(b). As a result, correlation functions along the diagonal directions [$(x,y)=(1, \pm 1)$] decay slightly faster than those along the axial directions [Fig.~\ref{fig:acf}(a)], with the corresponding correlation length being slightly ($\simeq 10$\%) smaller [Fig.~S2(b)]. However, this geometrical influence on coarsening dynamics is negligible, as discussed in Sec.~\ref{sec:tri}.
Owing o the shape change in domains, the correlation functions are not self-similar during the coarsening [compare the green and red curves in Fig.~\ref{fig:acf}(a)].
Overall, the LAC law is maintained at $q=3$, even under nonequilibrium conditions, except for a transient increase in $n_{\mathrm{cl}}^{1/2}$.

\begin{figure}[tbh]
\includegraphics[]{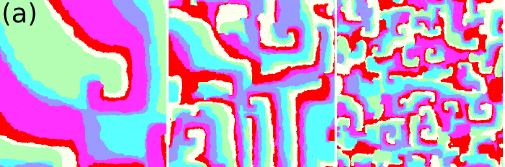}
\includegraphics[]{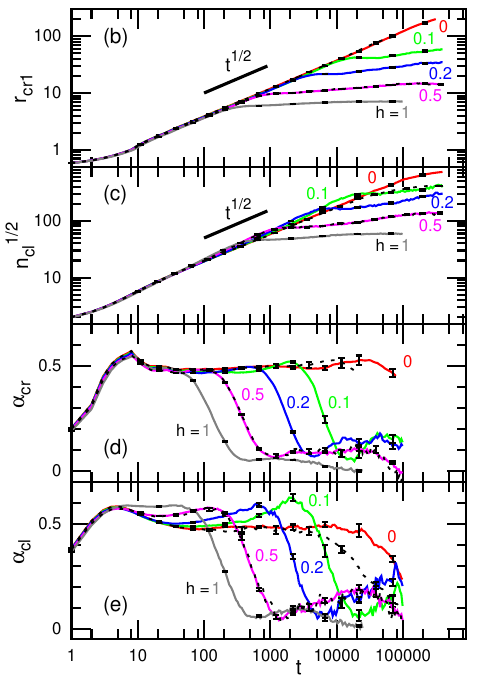}
\caption{
Coarsening dynamics of the six-state repulsive nonflip-contact Potts model ($J_{k,[k+2]}=J_{k,[k+3]}=-2$).
(a) Snapshots ($L/4\times L/4$) at $t=400~000$ for $h=0.2$,  $0.5$, and $1$ (left to right) and $L=2048$.
(b)--(e) Time evolution of (b) the correlation length $r_{\mathrm{cr1}}$, (c) root of the mean cluster size $n_{\mathrm{cl}}^{1/2}$, 
(d) exponents $\alpha_{\mathrm{cr}}$, and (e) $\alpha_{\mathrm{cl}}$.
The solid lines represent the data for $h=0$, $0.1$, $0.2$, $0.5$, and $1$ at $L=2048$.
The black dashed lines represent the data for $h=0$ and $0.5$ at $L=1024$.
The short black lines in (b) and (c) represent the slope of $t^{1/2}$.
The error bars are shown at several data points.
}
\label{fig:q6d2}
\end{figure}

\begin{figure}[tbh]
\includegraphics[]{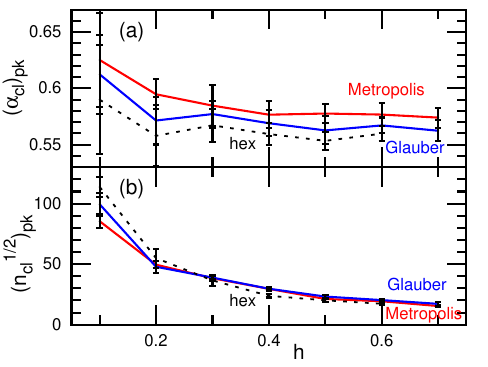}
\caption{
Peaks in coarsening exponents $\alpha_{\mathrm{cl}}$ as functions of $h$ for the six-state repulsive nonflip-contact Potts model.
(a) Peak heights of $\alpha_{\mathrm{cl}}$. (b) Cluster length $n_{\mathrm{cl}}^{1/2}$ at the peaks. 
The red and blue colors represent the data obtained using the Metropolis and Glauber update schemes, respectively, on the square lattice.
The black dashed lines represent the data obtained using the Metropolis scheme on the hexagonal lattice.
}
\label{fig:slope_q6d2}
\end{figure}

\begin{figure}[tbh]
\includegraphics[]{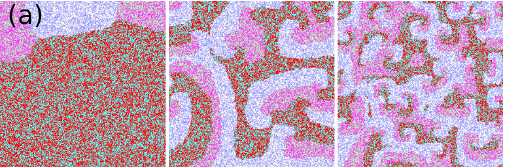}
\includegraphics[]{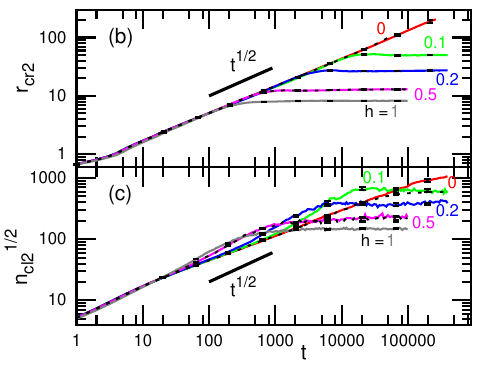}
\caption{
Coarsening dynamics of the six-state attractive diagonal-contact Potts model ($J_{k,[k+2]}=0$ and $J_{k,[k+3]}=1.5$).
(a) Snapshots ($L/4\times L/4$) at $t=100~000$ for $h=0$,  $0.2$, and $0.5$ (left to right) and $L=2048$.
Diagonal states [$s=0$ and $3$ (light yellow and blue); $s=1$ and $4$
(light green and magenta); and $s=2$ and $5$ (cyan and red), respectively]
are mixed, similar to those in a disorder phase. These three types of mixed domains form spiral waves
at $h>0$.
(b)--(c) Time evolution of (b) the correlation length $r_{\mathrm{cr2}}$ and (c) root of the mean cluster size $n_{\mathrm{cl2}}^{1/2}$
for mixed diagonal states ($s=k$ and $[k+3]$).
The solid lines represent the data for $h=0$, $0.1$, $0.2$, $0.5$, and $1$ at $L=2048$.
The black dashed lines represent the data for $h=0$ and $0.5$ at $L=1024$.
The short black lines represent the slope of $t^{1/2}$.
The error bars are shown at several data points.
}
\label{fig:q6tn15}
\end{figure}

\begin{figure}[tbh]
\includegraphics[]{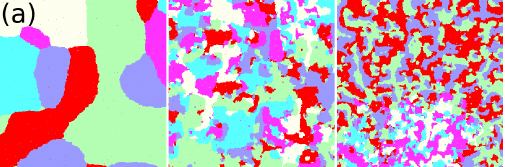}
\includegraphics[]{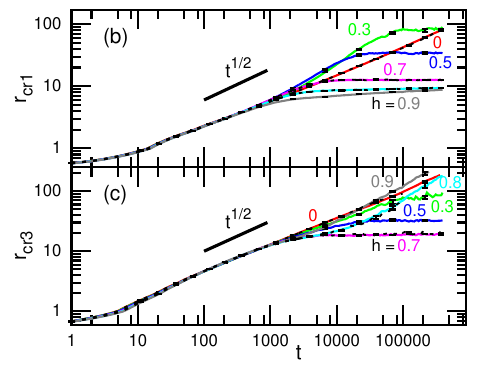}
\caption{
Coarsening dynamics of the six-state repulsive diagonal-contact Potts model ($J_{k,[k+2]}=0$ and $J_{k,[k+3]}=-1$).
(a) Snapshots ($L/4\times L/4$) at $t=400~000$ and $L=2048$ for $h=0$,  $0.7$, and $0.9$ (left to right).
In the right snapshot, the spiral waves of odd- and even-numbered states are formed in the upper and lower regions, respectively.
(b)--(c) Time evolution of the correlation lengths (b) $r_{\mathrm{cr1}}$ and (c) $r_{\mathrm{cr3}}$ for single and three (odd and even numbers) states, respectively.  
The solid lines represent the data for $h=0$, $0.3$, $0.5$, $0.7$, $0.8$, and $0.9$ at $L=2048$.
The black dashed lines represent the data for $h=0$, $0.7$, and $0.8$ at $L=1024$.
The short black lines represent the slope of $t^{1/2}$.
The error bars are shown at several data points.
}
\label{fig:q6t1}
\end{figure}

\subsection{Standard Potts Models}\label{sec:d0}

Next, we describe the coarsening dynamics of the standard Potts models at $q=4$--$8$ (Figs.~\ref{fig:q6d0}--\ref{fig:peak}, and S3--S7).
During coarsening to equilibrium states ($h=0$),
$r_{\mathrm{cr1}}$ and  $n_{\mathrm{cl}}^{1/2}$ grow as $t^{1/2}$ at $q=4$--$8$, similar to that at $q=3$
(see Figs.~\ref{fig:q6d0}, S4, S5, S6, and S7 for $q=6$, $4$, $5$, $7$, and $8$, respectively)~\cite{gres88}.
The coarsening exponent $\alpha_{\mathrm{cr}}$, averaged for $2<r_{\mathrm{cr1}}<4$, is lower with increasing $q$ 
compared with the exponent averaged for $30<r_{\mathrm{cr1}}<80$
[see red solid and black dashed lines in Fig.~\ref{fig:slope}(a)].
Hence, the approach to the asymptotic value of $1/2$ becomes slower with increasing $q$.

At $h>0$, nonspiral disordered waves emerge, and the domains remain approximately circular (see Figs.~\ref{fig:q6d0}(a), S4(a), and S7(a) for $q=6$, $4$, and $8$, respectively)~\cite{nogu25,nogu25b}.
Since flips do not occur at the boundaries between the domains of $s=k$ and $[k+n]$ with $2\le n\le q/2$,
the ballistic motion of domain boundaries occurs only between the $s=k$ and $[k+1]$ domains; therefore, the waves do not exhibit spiral shapes.
The correlation functions are isotropic [the dashed line of $C_{\mathrm{crd}}$ overlaps with the red line of $C_{\mathrm{cr1}}$ in Fig.~\ref{fig:acf}(b)]. 
In addition to  $n_{\mathrm{cl}}^{1/2}$,  $r_{\mathrm{cr1}}$ exhibits growth enhancement before saturation, unlike  $r_{\mathrm{cr1}}$ at $q=3$ (see Figs.~\ref{fig:q6d0}, S4, S5, S6, and S7 for $q=6$, $4$, $5$, $7$, and $8$, respectively).
The peak heights of the exponents $\alpha_{cr}$ and $\alpha_{cl}$ before the saturation increase with $q$, as shown in Fig.~\ref{fig:peak}(a).
The correlation length $r_{\mathrm{cr1}}$ at the peak increases with $q$ but the cluster length $n_{\mathrm{cl}}^{1/2}$ is almost independent [see Fig.~\ref{fig:peak}(b) and (c)].
The exponent $\alpha_{\mathrm{cr}}$ averaged for the time range before saturation increases
with increasing $h$ [see Fig.~\ref{fig:slope}(a)].

For the Pots models at $q=4$ and $6$, multiple contact types exist: $s=k$--$[k+1]$  and $k$--$[k+2]$  pairs at $q=4$,
and $k$--$[k+1]$,  $k$--$[k+2]$,  and $k$--$[k+3]$ pairs at $q=6$.
Since the boundaries between $s=k$ and $[k+1]$ domains move ballistically,
the contact probabilities $p_{\mathrm{cn1}}$ for $k$--$[k+1]$ pairs have higher values at $t=\infty$ than the other contact probabilities ($p_{\mathrm{cn2}}$ at $q=4$ and $p_{\mathrm{cn2}}$ and $p_{\mathrm{cn3}}$ at $q=6$), as shown in Figs.~S4(d), (e) and S3, respectively.
During saturation, the decrease in $p_{\mathrm{cn1}}$ is reduced, whereas the decreases in $p_{\mathrm{cn2}}$ and $p_{\mathrm{cn3}}$ are enhanced.

\subsection{Repulsive Nonflip-Contact Potts Models}\label{sec:d2}

Next, we describe the coarsening dynamics of repulsive nonflip-contact Potts models ($J_{k,[k+n]}=-J_{k,k}$ with $2\le n\le q/2$) 
at $q=4$--$8$ [Figs.~\ref{fig:slope}--\ref{fig:slope_q6d2}, and S8--S12].
Because nonflip contacts ($s=k$ and $[k+n]$ pairs with $2\le n\le q/2$) are suppressed by the repulsion,
spiral waves emerge at $h>0$  [see Figs.~\ref{fig:q6d2}(a), S9(a), and S12(a) for $q=6$, $4$, and $8$, respectively]~\cite{nogu25b}.

Interestingly, growth enhancement before the saturation is significantly reduced compared with that in the standard Potts models
(see Figs.~\ref{fig:q6d2}, S9, S10, S11, and S12 for $q=6$, $4$, $5$, $7$, and $8$, respectively),
and the exponents $\alpha_{\mathrm{cr}}$ and $\alpha_{\mathrm{cl}}$ at $q=6$ exhibit similar dependence to those at $q=3$.
Clear deviations from the $h=0$ slopes are not detected for $r_{\mathrm{cr1}}$ at $q=4$--$6$ and for $n_{\mathrm{cl}}^{1/2}$ at $q=4$ and $5$.
Subsequently, the growths gradually increase with increasing $q$,
so that the exponent $\alpha_{\mathrm{cl}}$ for the time range before the saturation increases
with increasing $q$ [see Figs.~\ref{fig:slope}(d) and \ref{fig:peak}(a)].
The peak heights decrease with increasing $h$ (see Fig.~\ref{fig:slope_q6d2}).
The contact probability of neighboring states $p_{\mathrm{cn1}}$ deceases as $t^{-1/2}$ 
but that of the second neighboring states $p_{\mathrm{cn2}}$ decreases rapidly (close to $t^{-1}$)
to reduce the contact energy [see Figs.~S8 and S9(d), (e)].
The reduction in these nonflip contacts
likely induces the reduction of the transient growth enhancement, although the domains have markedly more elongated shapes than those in the standard models.

\subsection{Attractive Diagonal-Contact Potts Model}\label{sec:q6tn15}

At $q=6$, C2 and C3 symmetric structures can emerge through symmetry factorization~\cite{nogu25b}, as in equilibrium systems~\cite{taka20,ditz80,gres81}.
When the diagonal contact is sufficiently attractive ($J_{k,[k+3]} =1.5$),
the diagonal states ($s=k$ and $[k+3]$) are mixed instead of undergoing phase separation.
Three types of mixed domains can then be formed: $s=0$ and $3$; $s=1$ and $4$; and $s=2$ and $5$.
At $h>0$, these three domains form spiral waves [see Fig.~\ref{fig:q6tn15}(a)], the dynamics of which is called the M2W3 mode~\cite{nogu25b}.
The coarsening dynamics of these mixed domains are almost identical to those of the three-state Potts model
(compare Figs.~\ref{fig:q6tn15} and S13(a), (b) with Fig.~\ref{fig:q3}).
During coarsening, the mixing of diagonal states first increases at $t \lesssim 100$,
and then slightly decreases [see Fig.~S13(c)].

\subsection{Repulsive Diagonal-Contact Potts Model}\label{sec:q6t1}

\begin{figure}[tbh]
\includegraphics[]{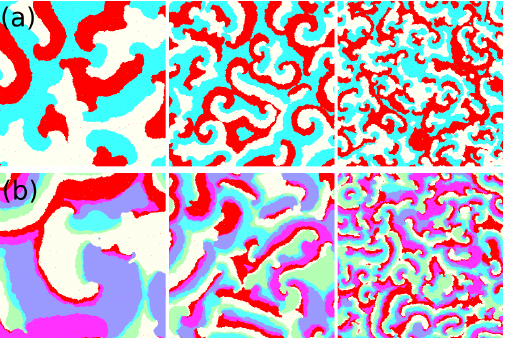}
\caption{
Snapshots ($L/4\times L/4$) of coarsening dynamics on the hexagonal lattice at $t=100~000$ and $L=2048$.
(a) $q=3$. (b) $q=6$ and repulsive diagonal-contact energy.
From left to right, $h=0.2$,  $0.5$, and $1$.
}
\label{fig:snaptri}
\end{figure}

When the diagonal contact is sufficiently repulsive ($J_{k,[k+3]} =-1$) at $q=6$,
the contact of second-neighboring states ($s=k$ and $[k+2]$) becomes relatively attractive; therefore,
 the odd- or even-numbered states form spiral waves at $h \ge 0.8$ [see the right snapshot in Fig.~\ref{fig:q6t1}(a)], which is called the W3 mode~\cite{nogu25b}.
In the spiral waves of odd-numbered states ($s=1$, $3$, and $5$), even-numbered states ($s=0$, $2$, and $4$) exist to a small degree at the domain boundaries.
The two spiral-wave regions of odd- and even-numbered states remain phase-separated.
Conversely, at $0<h \le 0.7$, all six states form disordered waves, as in the standard Potts model [see the middle snapshot in Fig.~\ref{fig:q6t1}(a)].

The correction length $r_{\mathrm{cr1}}$ of single states becomes saturated at a large $t$ for $h>0$, and 
the saturation length monotonically decreases with increasing $h$ [Fig.~\ref{fig:q6t1}(b)].
By contrast, the correction length $r_{\mathrm{cr3}}$ of the three odd- or even-numbered states
continuously increases for $h \ge 0.8$, whereas $r_{\mathrm{cr3}}$ becomes saturated for $0<h\le 0.7$ [Fig.~\ref{fig:q6t1}(c)].
Therefore, at $h=0.9$, the single-state correlation length $r_{\mathrm{cr1}}$ grows similarly to that of the spiral wave mode,
but  $r_{\mathrm{cr3}}$ grows similarly to that in spinodal decomposition for $h=0$.
The contact of second-neighboring states increases with increasing $h$,
whereas the contact of neighboring states ($k$ and $[k+1]$) reaches the maximum at $h=0.7$ (see Fig.~S14).

\subsection{Robustness of Coarsening Dynamics}\label{sec:tri}

We simulated coarsening on the hexagonal lattice and using the Glauber update scheme to examine the robustness of coarsening dynamics.
It is known that the coarsening dynamics of Potts models at zero temperature 
depend on lattice type, with
higher trap ratios into metastable states for square lattices~\cite{olej13} than for hexagonal lattices~\cite{denh21}.
In the present simulations of square lattices, spiral waves exhibit rectangular shapes elongating in the vertical and horizontal directions [$(x,y)=(0,\pm 1)$ and $(\pm 1, 0)$].
To examine the effects of this anisotropic domain structure,
simulations of Potts models are performed on hexagonal lattices at $q=3$ and $6$.
In these cases, isotropic spiral waves emerge, as shown in Fig.~\ref{fig:snaptri}.
Nevertheless, coarsening dynamics closely match those observed on square lattices (compare Figs.~S15 and S16 with Figs.~\ref{fig:q3} and \ref{fig:q6d2}, respectively),
with small differences in the obtained exponent values (see Figs.~\ref{fig:slope_q3} and \ref{fig:slope_q6d2}).
Therefore, we conclude that the influence of lattice structures on coarsening dynamics is negligible.

In this study, we use the Metropolis MC update scheme.
Since the Glauber update scheme  is also widely used for coarsening dynamics,
we also simulated the three- and six-state Potts models using the Glauber update.
 The resulting, coarsening dynamics are similar (Figs.~S17 and S18), and the exponent values calculated prior to saturation are slightly smaller than those calculated using the Metropolis scheme (Figs.~\ref{fig:slope_q3} and \ref{fig:slope_q6d2}). Since the acceptance ratio of the Glauber scheme is lower than that of the Metropolis scheme,
the approach to the asymptotic value is slightly slower.
Therefore, the choice of the update schemes does not result in a significant difference in the coarsening dynamics.

\section{Summary}\label{sec:sum}

We studied the coarsening dynamics of $q$-state active Potts models with $q=3$--$8$ under cyclically symmetric conditions. 
For equilibrium states ($h=0$), coarsening dynamics follow the LAC law (the characteristic length $\propto t^{1/2}$)
as reported in previous studies.
When active flips ($h>0$) are introduced into the three-state Potts model and repulsive nonflip-contact Potts models with $q=4$--$8$,
spiral waves emerge in the long-term limit.
By contrast, in the standard Potts models with $q=4$--$8$, the final states are nonspiral disordered waves (waves move separately without centers).
In both cases, the correlation length and the cluster length (the root of the mean cluster size) exhibit saturation to the characteristic wave lengths.
In the intermediate time ranges, these two lengths follow the LAC law
but exhibit a growth enhancement prior to saturation.
These growth enhancements are greater at higher $q$
and also greater in the cluster length than in the correlation length.
Therefore, the differences in the final spatiotemporal patterns modify the coarsening dynamics near saturation.

When a factorizable number of states is used for the Potts models, a mode of the factorized symmetry can emerge.
In this study, we examined two types of factorized modes at $q=6$: spiral waves composed of three types of diagonal mixed domains,
and the coexistence of spiral waves comprising odd- and even-numbered states.
In the former case, large domains comprising two states exhibit the  coarsening dynamics characteristic of $q=3$ spiral waves.
In the latter cases, large domains comprising three states exhibit the  coarsening dynamics characteristic of two-phase spinodal decomposition,
and a transient pattern appears during coarsening at the intermediate condition.
This type of transient patterns can be more clearly observed in the coarsening to stripe waves~\cite{nogu26c}.

Finally, this study examined the influence of the choice of lattice types (square vs. hexagonal lattices) and MC update schemes (Metropolis vs. Glauber).
Neither choice alters the  coarsening dynamic behavior, demonstrating that the obtained dynamics are robust with respect to model and algorithmic details.

\begin{acknowledgments}
The simulations were partially carried out at ISSP Supercomputer Center, University of Tokyo (ISSPkyodo-SC-2025-Ca-0049).
This work was supported by JSPS KAKENHI Grant Number JP24K06973. 
\end{acknowledgments}


%

\end{document}